\documentclass[showpacs,pre,twocolumn,floatfix]{revtex4}
\usepackage{amsmath}
\usepackage{graphicx}
\usepackage{bm}
\usepackage{CJK}     
\usepackage{ulem}   
\usepackage{color}  
\begin{document}
\begin{CJK*}{UTF8}{bsmi}
\title{Thomson scattering in the  average-atom approximation }
\author{W. R. Johnson}
\email{johnson@nd.edu}
\affiliation{Department of Physics, 225 Nieuwland Science Hall\\
University of Notre Dame, Notre Dame, IN 46556}
\author{J. Nilsen}
\author{K. T. Cheng (鄭國錚)}
\affiliation{Lawrence Livermore National Laboratory,
Livermore CA 94551}

\begin{abstract}
The average-atom model is applied to study Thomson scattering of x-rays from
warm-dense matter with emphasis on scattering by bound electrons.
Parameters needed to evaluate the dynamic structure function
(chemical potential, average ionic charge, free electron density,
bound and continuum wave functions and occupation numbers) are obtained from the
average-atom model. The resulting analysis provides a relatively simple diagnostic
for use in connection with x-ray scattering measurements.
Applications are given to dense hydrogen, beryllium, aluminum and titanium plasmas.
In the case of titanium, bound states are predicted to modify the
spectrum significantly.
\end{abstract}


\maketitle

\end{CJK*}

\section{Introduction}

Measurements of Thomson scattering of x-rays provide information on
temperatures, densities and ionization balance in warm dense matter.
Various techniques for inferring  plasma properties from x-ray scattering
measurements have been developed over the past decade
\cite{GGL:02,LB:02,GG:03,LM:03,GGL:03,GGL:03a,HR:04,GGL:06,GL:07,TB:08,KN:08,
FT:09,DL:09,KG:09,GL:10,MM:10,TB:10,RR:10,KD:11,VD:12,FL:12,ZB:12};
these techniques together with the underlying theory were reviewed
by \citet{GR:09}

The present analysis of Thomson scattering from dense plasmas is based on a
theoretical model proposed by \citet{GG:03}, one important difference being
that the parameters used here to evaluate the Thomson-scattering dynamic
structure function are taken from the average-atom model. The particular
average-atom model used here is described in Ref.~\cite{JGB:06}.
The present work is closely related to that of \citet{SG:08},
where a somewhat different version of the average-atom model was used.
Predictions from the present model differ substantially from those
in Ref.~\cite{SG:08}. The origin and consequences of these differences
will be  discussed later.

The Thomson scattering cross section for an incident photon with energy,
momentum ($\omega_0, \, {\bm k}_0$) and polarization ${\bm \epsilon}_0$
scattering to a state with energy,  momentum ($\omega_1,\,  {\bm k}_1$)
and polarization ${\bm \epsilon}_1$ is
\begin{equation}
 \frac{d\sigma}{d\omega_1 d\Omega} = \left(  \frac{d\sigma}{d\Omega}
   \right)_{\!\text{Th}}\frac{\omega_1}{\omega_0}\ S(k,\omega), \label{eq1}
\end{equation}
where
\begin{equation}
 \left(  \frac{d\sigma}{d\Omega} \right)_{\!\text{Th}} = \,
 |{\bm \epsilon}_0\cdot{\bm \epsilon}_1|^2 \left( \frac{e^2}{mc^2}
 \right)^{\! 2}.
\end{equation}
The {\it dynamic structure function} $S(k,\omega)$ appearing in Eq.~(\ref{eq1})
depends on two variables: $k = |{\bm k}_0 - {\bm k}_1|$ and
$\omega=\omega_0 - \omega_1$.
As shown in the seminal work of \citet{CH:87,CH:00}, $S(k,\omega)$ can be
decomposed into three terms: the first $S_{ii}(k,\omega)$ is the contribution
from elastic scattering by electrons that follow the ion motion,
the second $S_{ee}(k,\omega)$ is the contribution from scattering by
free electrons and the third  $S_B(k,\omega)$ is the contribution from
bound-free transitions (inelastic scattering by bound electrons) modulated
by the ionic motion. In the present work, the modulation factor is ignored
when evaluating the bound-free scattering structure function.
For the bound-free contribution, calculations carried out using plane-wave
final states are compared with calculations carried out using average-atom
scattering wave functions. Substantial differences are found between these
cases.

The average-atom model is discussed briefly in Sec.~\ref{avat} followed by
a discussion of the three contributions to the structure functions in
Sec.~\ref{dynam}. In Sec.~\ref{apps} applications are given to hydrogen,
beryllium, aluminum and titanium plasmas.

\section{Average-Atom Model\label{avat}}

Average-atom models are versions of the
temperature-dependent Thomas-Fermi model of a plasma developed
63 years ago by \citet{FMT:49}
which include detailed descriptions of bound and continuum states of
an atom imbedded in a plasma.
In this model, the plasma is divided into neutral Wigner-Seitz (WS) cells
(volume per atom $V_\text{\tiny WS} = A/\rho N_{\rm A}$,
where $A$ is the atomic weight, $\rho$ is the mass density and
 $N_{\rm A}$ is Avogadro's number).
Inside each WS  cell is a nucleus of charge $Z$ and $Z$ electrons.
Some of these electrons are in bound states and some in continuum states.
The continuum density is finite at the cell boundary and merges into the
uniform free-electron density
$Z_f/V_\text{\tiny WS}$
outside the cell. Each neutral cell can, therefore, be regarded
as an ion imbedded in a uniform sea of free electrons of density
$n_e = Z_f/V_\text{\tiny WS}$.
To maintain overall neutrality, it is necessary to introduce a uniform
(but \mbox{inert}) positive background density $Z_f/V_\text{\tiny WS}$.
The model, therefore, describes an isolated (neutral) ion floating in a
(neutral) ``jellium'' sea.

The quantum-mechanical model here, which is discussed in Ref.~\cite{JGB:06},
is a nonrelativistic version of the relativistic {\it Inferno} model of
\citet{DL:79} and the more recent {\it Purgatorio} model of \citet{BS:06};
it is similar to the nonrelativistic average-atom model described
by \citet{BI:95}. Specifically, each electron in the ion is assumed
to satisfy the central-field Schr\"{o}dinger equation
\begin{equation}
 \left[ \frac{p^2}{2} -\frac{Z}{r} + V \right] \psi_a({\bm r})
= \epsilon_a\, \psi_a({\bm r}) , \label{sq1}
\end{equation}
where $a=(n,l)$ for bound states or $(\epsilon,l)$ for continuum states.
Atomic units (a.u.) where
$e=\hbar=m=4\pi\epsilon_0=1$ are used here.
In particular, 1 a.u.\ in energy equals 2 Rydbergs (27.211 eV),
and 1 a.u.\ in length equals 1 Bohr radius $a_0$ (0.529 \AA).
The wave function $\psi_a({\bm r})$ is decomposed in a spherical basis as
\begin{equation}
  \psi_a({\bm r}) = \frac{1}{r} P_a(r)\, Y_{l_am_a}(\hat{r})\, \chi_{\sigma_a},
\end{equation}
where $Y_{lm}(\hat{r})$ is a spherical harmonic and
$\chi_{\sigma}$ is a two-component electron spinor.
The bound and continuum radial functions $P_a(r)$ are normalized as
\begin{align}
\int_0^\infty\!\!  dr P_{nl}(r)\, P_{n'l}(r) &={} \delta_{nn'}, \\
\int_0^\infty\!\!  dr P_{\epsilon l}(r)\, P_{\epsilon' l}(r) &={}
\delta(\epsilon-\epsilon') ,
\end{align}
respectively. The central potential $V(r)$ in Eq.~(\ref{eq1})
is taken to be the self-consistent Kohn-Sham potential \cite{kohnsham}
\begin{equation}
V(r) = 4\pi \!\int\! \frac{1}{r_>} \, r'^2 \, n(r') \, dr'
- x_\alpha \bigg[ \frac{81}{8\pi} \, n(r) \bigg]^{\!\frac{1}{3}},
\label{sq5}
\end{equation}
where the first term in the right-hand side of Eq.\ (\ref{sq5})
is the direct screening potential with $r_> = \text{max}(r,r')$
and the second term is the average exchange potential with $x_\alpha=2/3$.
While short-range electron-electron
interactions inside the Wigner-Seitz cells are reasonably well
accounted for by this simple model, it should be noted that
eigenvalues in the Kohn-Sham potential are poor approximations to
ionization energies,
leading to inaccurate thresholds and peaks of bound-free contributions
to $S(k,\omega)$, which can be off
by 20 -- 30\% when compared with experiment.

The electron density $n(r)$ in Eq.~(\ref{sq5}) has contributions from
bound-states $n_b(r)$ and from continuum states $n_c(r)$,
\begin{equation}
 n(r) = n_b(r)+n_c(r).
 \end{equation}
The bound-state contribution to the density $n_b(r)$ is
\begin{equation}
  4\pi r^2 n_b(r) = \sum_{nl} \frac{2(2l+1)}
  {1+ \exp[(\epsilon_{nl}-\mu)/k_{\scriptscriptstyle B}T]}\,
   P_{nl}(r)^2 , \label{sq3}
\end{equation}
where $\epsilon_{nl}$ is the bound-state energy, $\mu$ is the chemical
potential, and the sum over $(n,l)$ ranges over all bound subshells.
The continuum contribution to the density $n_c(r)$ is given by
a similar expression with the bound state radial functions $P_{nl}(r)$
replaced by continuum functions $P_{\epsilon l}(r)$ and the sum over $n$
replaced by an integral over $\epsilon$.
Finally, the chemical potential $\mu$ is chosen to ensure charge neutrality
inside the WS cell:
\begin{equation}
Z = \int_{r \le R_\text{\tiny WS}}\!\! n(r)\, d^3r\equiv
\int_0^{R_\text{\tiny WS}} \!\! 4\pi r^2 n(r)\, dr \ . \label{sq6}
\end{equation}
Equations (\ref{sq1}-\ref{sq6}) above are solved self-consistently
to give the chemical potential $\mu$, the potential $V(r)$ and the
electron density $n(r)$.
Numerical details can be found in Ref.~\cite{JGB:06}.

\begin{figure}[t]
\centerline{\includegraphics[scale=0.7]{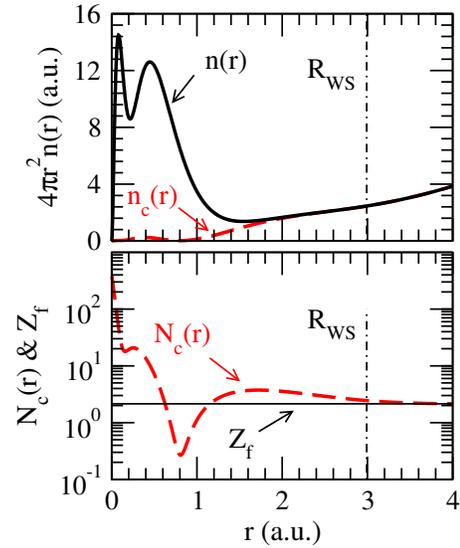}}  
\caption{(Color online) Upper panel: The radial density $4\pi r^2 n(r)$
for Al at metallic density and
$k_{\scriptscriptstyle B} T=5$~eV (solid curve)
integrates to $Z=13$ for $r\leq R_\text{\tiny WS}$. The continuum
contribution $4\pi r^2 n_c(r)$ (dashed curve) integrates to 3 for
$r\leq R_\text{\tiny WS}$. The bound $1s$, $2s$ and $2p$ shells
are completely occupied at this temperature.
Lower panel: The dashed curve illustrates the Friedel oscillations of
the continuum density and shows how $N_c(r) = n_c(r) V_{\text{\tiny WS}} $
converges to $Z_f = n_e V_{\text{\tiny WS}}$ (solid line) for
$r>R_\text{\tiny WS}$. The chemical potential predicted by the model is
$\mu = 0.2406$~a.u.\ and the number of free electrons per ion is $Z_f = 2.146$.
\label{figg}}
\end{figure}

The boundary conditions used in solving Eq.~(\ref{sq1}) deserve some mention.
Bound state wave functions and their derivatives are matched at the boundary
$r=R_\text{\tiny WS}$ to solutions outside the WS sphere (where $V=0$)
that vanish exponentially as $r\to \infty$.
Similarly, continuum functions and their derivatives are matched to
phase-shifted free-particle wave functions at $r=R_\text{\tiny WS}$.
It should be noticed that the continuum density $n_c(r)$ inside the WS sphere,
which oscillates as predicted by \citet{JF:54},
is distinctly different from the uniform free electron density $n_e$.
In the present model, $n_c(r)$ smoothly approaches $n_e$ outside the sphere.
These points are illustrated in Fig.~\ref{figg}, where the bound-state and
continuum densities are plotted for Al at metallic
density and temperature $k_{\scriptscriptstyle B} T = 5$~eV.

The boundary conditions used here differ from those used
by Sahoo et al.\ in Ref.~\cite{SG:08},
where the first derivative of the wave function is required to vanish at
$R_\text{\tiny WS}$. The differences in boundary conditions lead to major
differences in the average-atom structure. For example, the model used
in \cite{SG:08} predicts that the $M$ shell of
Al at metallic density is partially
occupied at temperatures $k_{\scriptscriptstyle B} T\leq 10$~eV,
whereas the present model predicts that the $M$ shell is empty in this
temperature range.
Consequences of such differences are discussed later in Sec.~\ref{apps}.

\section{Dynamic Structure Function\label{dynam}}

In the paragraphs below, the evaluation of $S(k,\omega)$ in the average-atom
approximation is discussed. As mentioned earlier, the theoretical model
developed by \citet{GG:03} is used to evaluate the ion-ion contribution
$S_{ii}(k,\omega)$ to the dynamic structure function. Additionally, the
procedure proposed in Ref.~\cite{GGL:06} is used to account for differences
between electron and ion temperatures. The electron-electron contribution
$S_{ee}(k,\omega)$ is expressed in terms of the dielectric function
$\epsilon(k,\omega)$ of the free electrons which in turn is evaluated
using the random-phase approximation (RPA). Finally, bound-state contributions
to the dynamic structure function are evaluated using average-atom bound state
wave functions. The final-state wave function is described in two different
ways: (1) using a plane-wave final-state wave function as in Ref.~\cite{SG:08},
and (2) using an average-atom final-state wave function that approaches a
plane wave asymptotically. There are dramatic differences between these
choices. The more realistic average-atom choice automatically includes
ionic Coulomb-field effects.

\subsection{Ion-Ion Structure Function\label{ion}}

The contribution to the dynamic structure function from elastic scattering by
electrons following the ion motion $S_{ii}(k,\omega)$ is expressed in terms of
the corresponding static ion-ion structure function $S_{ii}(k)$ as:
\begin{equation}
S_{ii}(k,\omega) = |f(k)+ q(k)|^2 \, S_{ii}(k)\, \delta(\omega) .
\label{eqsii}
\end{equation}
In the above, $f(k)$ is the Fourier transform of the bound-state density
and $q(k)$ is the Fourier transform of electrons that screen the ionic charge.
In the average-atom approximation, the screening electrons are the continuum
electrons inside the Wigner-Seitz sphere and
\begin{equation}
f(k)+q(k) = 4 \pi \!\int_0^{R_{\scriptscriptstyle WS}} \!\! r^2
\left[ n_b(r)+n_c(r) \right] j_0(kr) dr ,
\end{equation}
where $j_l(z)$ are spherical Bessel functions of order $l$.
Note that $f(0)+q(0) = Z$ in the average-atom model.
Furthermore, the delta function $\delta(\omega)$ in Eq.\ (\ref{eqsii})
is replaced by an ``instrumental''  Gaussian,
with full-width at half maximum = 10 eV in this work.
This value is chosen because
typical experiments in Be \cite{DL:09} have a spectrometer with a 10 eV
instrument width and use a Cl Ly-$\alpha$ source at 2.96 keV.

Approximate schemes to evaluate the static structure functions $S_{ii}(k)$
are discussed, for example, in Ref.~\cite{JPH:06}.
Here, we follow Ref.~\cite{GG:03} and make use of formulas given by \citet{AD:98}
that account for both quantum-mechanical and screening effects.
The function $S_{ii}(k)$ in Ref.~\cite{AD:98} is expressed in terms of the
Fourier transform of the ion-ion interaction potential $\Phi_{ii}(r)$
through the relation:
\begin{equation}
S_{ii}(k) = 1 - \frac{n_i}{k_{\scriptscriptstyle B} T} \Phi_{ii}(k),
\end{equation}
where $n_i$ is the ion density.
\begin{figure}
\centerline{\includegraphics[scale=0.7]{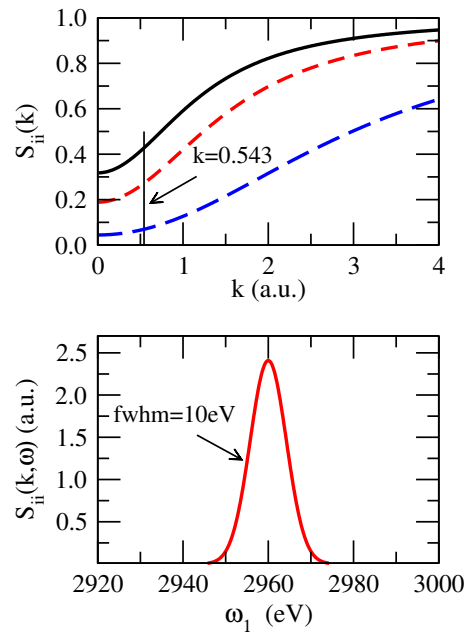}}
\caption{(Color online) Upper panel: $S_{ii}(k)$ is shown for Be metal
at electron temperature $T_e$=20 eV and ion-electron temperature ratios
$T_i/T_e$=(1,\, 0.5,\, 0.1) illustrated in solid, short dashed and
long dashed curves, respectively.
The value $k$ = 0.543 corresponds to an incident photon $\omega_0=2960$~eV
scattered at angle 40$^\circ$. Lower panel: $S_{ii}(k,\omega)$ for Be metal
at $T_e=20$ eV and $T_i=2$ eV, where the function $\delta(\omega)$ is
replaced by a Gaussian of width 10~eV and $k=0.543$.
\label{fig2}}
\end{figure}

\paragraph*{Different Electron and Ion Temperatures}
In the average atom model, $T$ is the electron temperature
$T_e$ which, in equilibrium, is equal to the ion temperature $T_i$.
To allow for different electron and ion temperatures, the equations for
$S_{ii}(k)$ given by \citet{AD:98} are modified following
the prescription laid out by \citet{GGL:06}.
The electron temperature $T_e$ is replaced by
an effective temperature $T_e^\prime$ that accounts for degeneracy effects
at temperatures lower than the Fermi temperature $T_F$. Similarly, the
ion temperature $T_i$ is replaced by an effective temperature $T_i^\prime$
that accounts for ion degeneracy effects at temperatures lower than the ion
screened Debye temperature $T_D$. Explicit formulas for $S_{ii}(k)$ are
found in Ref.~\cite{GGL:06}. The dramatic effect of different electron and ion
temperatures on the static structure functions $S_{ii}(k)$ for
Be at metallic density and
$T_e=20$~eV are illustrated in the top panel of Fig.~\ref{fig2}.
This figure is similar to the upper-left panel of Fig.~1 in Ref.~\cite{GGL:06},
which was obtained under similar condition.
In the bottom panel of Fig.~\ref{fig2} contributions to $S_{ii}(k,\omega)$
for Be at $T_e=20$~eV and  $T_i=2$~eV are shown.

\subsection{Electron-Electron Structure Function\label{elec}}
\begin{figure}
\centerline{\includegraphics[scale=0.65]{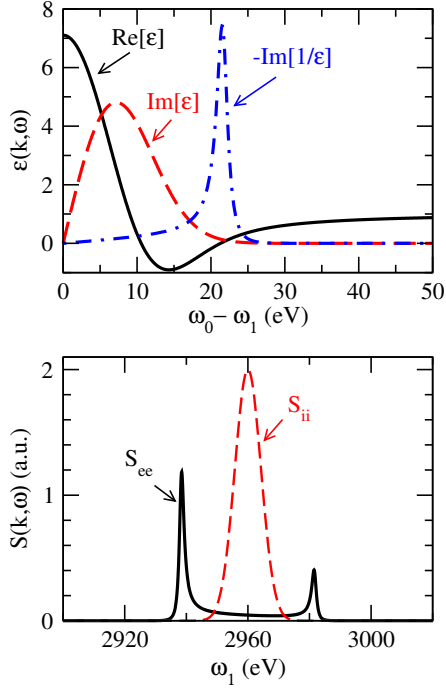}}  
\caption{(Color online) Upper panel: Real (solid) and imaginary (dashed)
parts of $\epsilon(k,\omega)$ are plotted along with
Im$[1/\epsilon(k,\omega)]$ (dot-dashed) for Be metal at
$k_{\scriptscriptstyle B}T = 20$ eV. Lower  panel:
The resulting structure function $S_{ee}(k,\omega)$ (solid) is shown together
with $S_{ii}(k,\omega)$  (dashed). These plots correspond to
Thomson scattering of a 2960eV photon at 20$^\circ$. \label{figeps}}
\end{figure}
The electron-electron structure function $S_{ee}(k,\omega)$ is expressed
in terms of the plasma dielectric function $\epsilon(k,\omega)$ through
Eq.\ (15) in Ref.~\cite{GG:03}:
\begin{equation}
S_{ee}(k,\omega) = 
- \frac{1}{1-\exp(-\omega/k_{\scriptscriptstyle B}T)}\,
\frac{k^2}{4\pi n_e}\,
{\rm Im}\!\left[\frac{1}{\epsilon(k,\omega)}\right] .
\end{equation}
In the average atom model, the free electrons are uniformly distributed
outside the WS sphere. The density of these electrons is
$n_e = Z_f/V_{\text{\tiny WS}}$. In the present work,
the dielectric function is evaluated using the random-phase approximation.
The real and imaginary parts of the RPA dielectric function
$\epsilon(k,\omega)$, given in Eq.~(16) of Ref.~\cite{GG:03},
can be written as
\begin{multline}
 {\rm Re} [ \epsilon(k,\omega) ] = 1 + \frac{2}{\pi k^3} \!\int_0^\infty\!\!
\mathcal{F}(p)\, p\, dp
\\ \times
 \left[ \ln \left|\frac{k^2+2pk+2\omega}{k^2-2pk+2\omega}\right| +
 \ln\left|\frac{k^2+2pk-2\omega}{k^2-2pk-2\omega}\right| \right]
\end{multline}
and
 \begin{multline}
 {\rm Im}[\epsilon(k,\omega)]
      = \frac{2}{k^3} \!\int_a^b\! \mathcal{F}(p)\,p\,dp
\\
      = \frac{2 k_{\scriptscriptstyle B} T}{k^3} \log\left[
        \frac{1+\exp[(\mu-a^2/2)/k_{\scriptscriptstyle B} T]}
       {1+\exp[(\mu-b^2/2)/k_{\scriptscriptstyle B} T]}\right]
\end{multline}
with $a = |2\omega-k^2|/2k$ and $b=(2\omega+k^2)/2k$. In these equations,
\begin{equation}
  \mathcal{F}(p) = \frac{1}{1+\exp[(p^2/2-\mu)
  /k_{\scriptscriptstyle B}T]}
\end{equation}
is the free-electron Fermi distribution function.
It should be noted that the real part of $\epsilon(k,\omega)$ is an even
function of $\omega$ and the imaginary part is an odd function of $\omega$.

The real and imaginary parts of $\epsilon(k,\omega)$
along with $-{\rm Im}[1/\epsilon(k,\omega)]$ are illustrated
in the top panel of Fig.~\ref{figeps} for scattering of a 2960 eV photon
at 20$^\circ$ from Be metal at 20 eV.
The sharp peak in ${\rm Im}[1/\epsilon]$ that occurs near the point where
${\rm Re}[\epsilon]$ vanishes is a collective plasma resonance (plasmon).
The contribution to $S_{ee}(k,\omega)$ is shown in the bottom panel.
The ratio of the down-shifted ($\omega_1 < 2960$ eV)
to up-shifted ($\omega_1 > 2960$ eV) resonance peaks
$\exp(\Delta \omega/{k_{\scriptscriptstyle B}T})$,
where $\Delta \omega$ is the energy of the plasmon peak
relative to the central energy, is used to determine
the electron temperature.

\subsection{Scattering from Bound States\label{boun}}
\begin{figure}
\centerline{\includegraphics[scale=0.45]{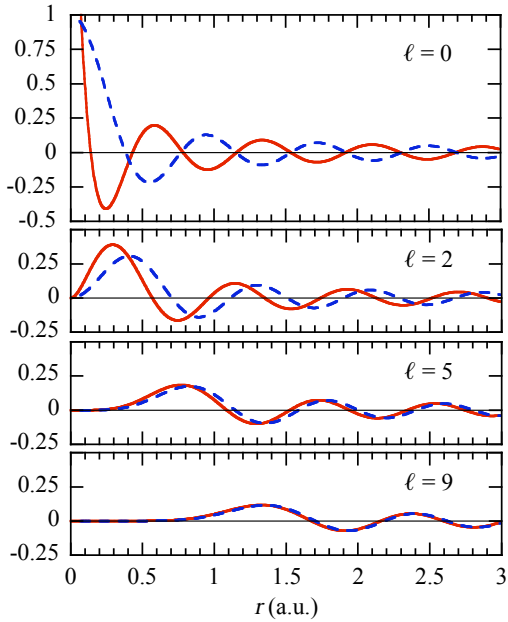}} 
\caption{(Color online) Comparison of average-atom continuum functions
$P_l(pr)/pr$ (solid lines) for Al at
metallic density and
$k_{\scriptscriptstyle B} T=5$~eV
with spherical Bessel functions $j_l(pr)$ (dashed lines).
\label{fig4}}
\end{figure}

The structure function associated with Thomson scattering from a bound state
$\psi_{nl}(\bm{r})$ with quantum numbers $(n,\, l)$ to a continuum state
$\psi_p(\bm{r})$ with momentum $\bm{p}$ is
\begin{equation}
S_{nl}(k,\omega) =
\sum_{m} \!\int\! \frac{ p\, d\Omega_p}{  (2\pi)^3 } \left|
 \int\! d^3r\, \psi^\dagger_{p}(\bm{r})\, e^{i \bm{k}\cdot\bm{r} }\,
 \psi_{nlm}(\bm{r})
\right|^2_{\epsilon_p=\omega+\epsilon_{nl}} \! .
\label{snl}
\end{equation}
As mentioned previously, two possibilities are considered for the final state
in bound-free scattering: (1) a free-particle plane wave, and
(2) an average-atom continuum wave that approaches a plane wave asymptotically.
Case (2) is clearly the more physical alternative since continuum waves
in the average-atom potential differ markedly from free-particle wave functions.
This point is illustrated in Fig.~\ref{fig4}, where the
average atom radial-functions $P_{\epsilon l}(r)/pr$ are compared with
their free-particle counterparts $j_l(pr)$. The average-atom wave functions
are seen to differ markedly from the free-particle (spherical Bessel)
functions for low values of $l$, but approach free-particle
functions as $l$ increases.

\paragraph{Plane-wave final states}

Assuming that the final state wave function is a free-particle
plane wave $e^{i \bm{p} \cdot \bm{r}}$,
the bound-free structure function in Eq.\ (\ref{snl}) can be rewritten as
\begin{equation}
S_{nl}(k,\omega) =
\sum_{m} \int \frac{ p\, d\Omega_p}{(2\pi)^3}
\left| \int\! d^3 r\, e^{i \bm{q}\cdot\bm{r}}\, \psi_{nlm}(\bm{r})
\right|^2_{\epsilon_p=\omega+\epsilon_{nl}}
 ,\label{sb}
\end{equation}
where ${\bm k}={\bm k}_0 - {\bm k}_1$, $\omega = \omega_0-\omega_1$ and
${\bm q}= {\bm k}-{\bm p}$.
Note that ${\bm q}$ is the momentum transferred to the ion. This expression
may be simplified to
\begin{equation}
 S_{nl}(k,\omega) = \frac{o_{nl}}{\pi k} \int_{|p-k|}^{p+k} q\, dq\,
 |K_{nl}(q)|^2, \label{fin}
\end{equation}
where $o_{nl}$ is the occupation number of the final state and
\begin{equation}
 K_{nl}(q) = \int_0^\infty\!\! dr\, r\,j_{l}(qr) P_{nl}(r) .
\end{equation}
Eq.~(\ref{fin}) depends implicitly on $\omega$  through the relation
\[
p = \sqrt{2(\omega+ \epsilon_{nl}})\ .
\]

\begin{figure}[t]
\centerline{\includegraphics[scale=0.6]{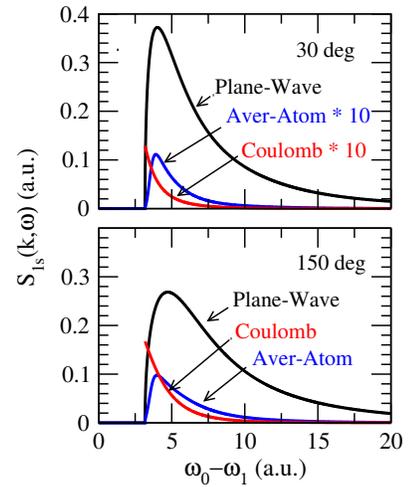}}  
\caption{(Color online) The beryllium $K$ shell structure function
$S_{1s}(k,\omega)$ is shown for incident photon energy 2960 eV and
scattering angles 30$^\circ$ and 150$^\circ$.
The black curves show plane-wave results,
the blue lines show the result obtained using an average atom
final-state wave function and the red lines show exact nonrelativistic
Coulomb results. The dramatic suppression of average atom and Coulomb
structure functions at forward angles (the corresponding curves
are multiplied by 10) is evident in the upper panel.\label{beaa}}
\end{figure}

\paragraph{Average-atom final states}
In the average-atom approach, the final state wave function
consists of a plane wave plus an {\it incoming}
spherical wave. (n.b.\ An outgoing spherical wave is
associated with an incident electron. Time-reversal invariance,
therefore, requires that a converging spherical wave be associated
with an emerging electron.)
The bound-free structure function in Eq.\ (\ref{snl}) may be reexpressed as
\begin{equation}
S_{nl} =  \frac{2p}{\pi}\, o_{nl}  \sum_{l_1l_2} A_{l_1l\,l_2} \,
|I_{l_1l\,l_2}(p,k) |^2,
\end{equation}
where $o_{nl}$ is the occupation number of the final state with
\begin{equation}
I_{l_1l\,l_2}(p,k) = \, \frac{1}{p}\,\, e^{i\delta_{l_1}(p)}
\!\int_0^{R_\text{\tiny WS}}\!\! dr\, P_{\epsilon l_1}(r)\, j_{l_2}(kr) P_{nl}(r).
\end{equation}
and
\begin{equation}
A_{l_1l\,l_2} = (2l_1+1)(2l_2+1) \left( \begin{array}{ccc}
l_1 & l & l_2 \\
0 & 0 & 0
\end{array} \right) .
\end{equation}
In the above, $\delta_{l_1}(p)$ is the phase-shift of the final state
partial wave $P_{\epsilon l_1}(r)$.
Moreover, $\epsilon = \omega+\epsilon_{nl}$, $p = \sqrt{2\epsilon}$ and
$k = |\bm{k}_0-\bm{k}_1|$.

In Fig.~\ref{beaa}, several calculations of the structure function
$S_{1s}(k,\omega)$ are compared for a photon of incident energy 2960~eV
scattered at 30$^\circ$ and 150$^\circ$ from the $K$ shell of beryllium metal
at $T$ = 20 eV. The results of calculations carried out using average-atom
final states are smaller than those using plane-wave final states by
a factor of about 40 at forward angles and 2.5 at backward angles.
This suppression is a characteristic Coulomb field effect. Indeed, exact
nonrelativistic Coulomb-field calculations of Thomson scattering \cite{EP:70},
with nuclear charge adjusted to align the Coulomb and average atom thresholds,
show a similar suppression.

\section{Applications\label{apps}}

In the subsections below, $S(k,\omega)$ is evaluated in the average-atom
approximation for cases of possible experimental interest:
hydrogen at $n_e = 10^{24}$~cm$^{-3}$ and $T=50$~eV,
beryllium at $n_e=1.8 \times 10^{23}$~cm$^{-3}$ and $T=18$~eV,
aluminum at metallic density and $T=5$~eV and
titanium at metallic density and $T=10$~eV.

\begin{figure}
\centerline{\includegraphics[scale=0.7]{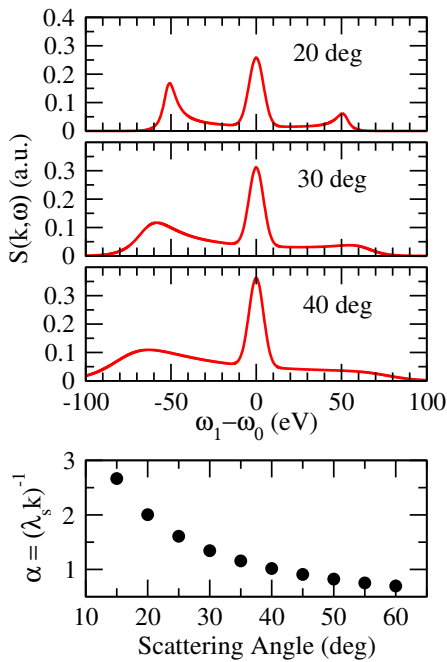}}
\caption{(Color online) The evolution of plasmon resonances
for scattering of a 5~keV photon from a fully ionized dense H plasma
($T$ = 50~eV, $n_e$ = 10$^{24}$ cm$^{-3}$)
is illustrated in the top panel where we plot $S(k,\omega)$ for scattering
angles of 20$^\circ$, 30$^\circ$ and 40$^\circ$. The corresponding coherence
parameters $\alpha = 1 / \lambda_s k$ are plotted
in the bottom panel. \label{hsfun} }
\end{figure}

\subsection{Hydrogen at $T= 50$~eV and $n_e = 10^{24}$~cm$^{-3}$}

In the average-atom model, a density $\rho$ =1.931 g/cc is required at
$T = 50$ eV to achieve free-electron density $n_e=10^{24}$~cm$^{-3}$.
The chemical potential in this case is $\mu=-1.091$~(a.u.).
Under these conditions of temperature and density, hydrogen is completely
ionized. The continuum density $n_c(r)$ inside the WS sphere merges into
the free-electron density $n_e$ outside the sphere. The total number of
electrons inside the WS sphere
$N_c =4\pi \int_0^{R_\text{\tiny WS}} r^2 n_c(r) dr =1$,
however, $Z_f = 0.8667$.

Since there are no bound electrons in this case, only $S_{ii}$ and $S_{ee}$
contribute to the cross section. Dynamic structure functions for scattering
of a 5~keV photon at angles ranging from 20$^\circ$, 30$^\circ$ and 40$^\circ$
are shown in the top panel of Fig.~\ref{hsfun}. Resonance peaks are seen
to broaden and move to higher frequencies as the scattering angle increases.
The coherence parameter $\alpha = 1/(\lambda_s k)$, defined in Eqs.~(5-7)
of Ref.~\cite{GR:09}, is plotted
in the bottom panel of Fig.~\ref{hsfun}.
The parameter $\lambda_s$ is the shielding length, given by
\begin{equation}
\lambda_s = \sqrt{
\frac{k_{\scriptscriptstyle B}T F_{1/2}(\mu/{k_{\scriptscriptstyle B}T})}
{4 \pi n_e F_{-1/2}(\mu/{k_{\scriptscriptstyle B}T})}} \, ,
\end{equation}
where $F_j(x)$ is a complete Fermi-Dirac integral,
\begin{equation}
F_\nu(x) = \frac{1}{\Gamma(1+\nu)} \int_0^\infty\! \frac{y^\nu}{1+\exp(y-x)} \, .
\end{equation}
For this particular case, $\lambda_s = 1.071$~a.u..
The value of $\lambda_s$ differs only slightly from
the WS radius $R_\text{\tiny WS}=1.118$ a.u..
The resonant features in Fig.~\ref{hsfun} are distinct for $\alpha>1$
but disappear for $\alpha \leq 1$, in harmony with the fact that
plasmon resonances are collective phenomena.
It should be noted that the (unperturbed) plasma frequency for
hydrogen at $n_e = 10^{24}$~cm$^{-3}$ is $\omega_\text{pl} = 37.1$~eV.

\subsection{Beryllium at $T = 18$~eV and $n_e = 1.8 \!\times\! 10^{23}$~cm$^{-3}$}

\begin{figure}
\centerline{\includegraphics[scale=0.7]{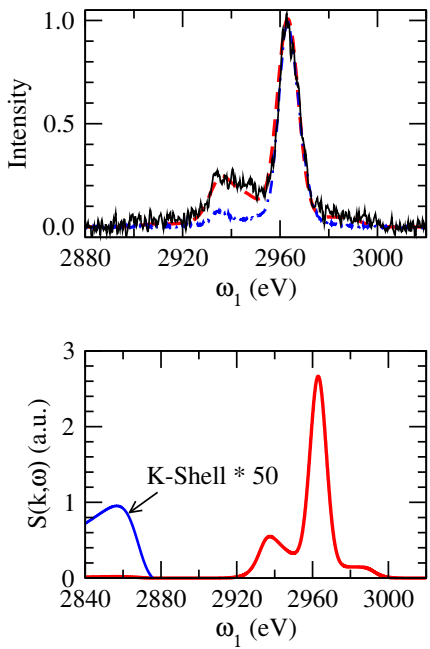}}
\caption{(Color online) Bottom panel: Structure function $S(k,\omega)$
for scattering of a 2963~eV photon at 40$^\circ$ from
beryllium at metallic density and
$T_e=18$~eV. Top panel: Intensity measurements for
scattering of a Cl Ly-$\alpha$ x-ray from beryllium at 40$^\circ$
\cite{GD:12}; measurement (solid line), source function (dot-dashed line),
and average atom fit (dashed line).
\label{be20}}
\end{figure}

In the bottom panel of Fig.~\ref{be20}, the structure function for
scattering of a 2963~eV photon
at 40$^\circ$ from beryllium (density = 1.636~g/cc) at $T_e=18$~eV
is plotted. The $L$ shell electrons are completely stripped under these
conditions but the $K$ shell remains 97\% occupied. The chemical potential
is $\mu = -0.5311$~a.u.\ and the number of free electrons per ion
$Z_f = 1.647$. The ion temperature, which governs the amplitude of
the elastic peak, is chosen to be $T_i= 2.1$~eV in this example.
For the case at hand, the coherence parameter is $\alpha = 1.21$,
so one expects and observes plasmon peaks in the scattering intensity profile.
The average-atom removal energy for a $K$ shell electron is 86.8~eV.
One therefore expects to find a contribution to 
$S(k,\omega)$ from $K$ shell
electrons for energies $\omega_1 < 2876$~eV.
The $K$ shell contribution multiplied by 50 is shown in the bottom panel.

To validate the present average-atom model against experimental data,
a Be experiment done at the Omega laser facility that used a Cl Ly-$\alpha$
source to scatter from nearly solid Be at an angle of 40$^\circ$ is used.
An electron temperature of 18~eV, ion temperature of~2.1 eV,
and density of 1.647~g/cc used in the average-atom model
gives an electron density of $1.8 \!\times\! 10^{23}$/cc,
in agreement with the analysis in Ref.~\cite{GD:12}.
The top panel of Fig.~\ref{be20} shows the experimental source function
from the Cl Ly-$\alpha$ line as a blue dashed line. Because of satellite
structure in the source we approximate the source by three lines:
a Cl Ly-$\alpha$ line at 2963~eV with amplitude 1 and two satellites
at 2934 and 2946~eV with relative amplitudes of 0.075 and 0.037 respectively.
Doing the Thomson scattering calculation using the three weighted lines,
we calculate the scattering amplitude for Thomson scattering (red dashed line)
and compare against the experimental data (black solid line) here.
We observe excellent agreement within the experimental noise.
Contributions from the bound $1s$ electrons, which have a threshold at 2876~eV,
are beyond the range of the data shown in the top panel.

\subsection{Aluminum at $T = 5$~eV and metallic density}

\begin{figure}[t]
\centerline{\includegraphics[scale=0.7]{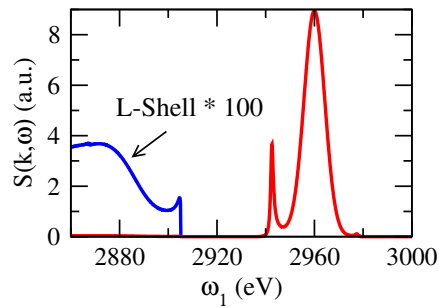}}  
\caption{(Color online) Structure function $S(k,\omega)$ for scattering
of a 2960~eV photon at 30$^\circ$ from metallic density
aluminum ($Z=13$) at $T_e=5$~eV and $T_i/T_e=0.1$. Contributions from the
$L$ shell
of the Ne-like core multiplied by 100 are indicated on the plot.
\label{al2}}
\end{figure}

Aluminum at metallic density
($\rho = 2.70$~gm/cc) and $T=5$~eV has a Ne-like ion
configuration with two $2s$ electrons bound by 92.2~eV and
six $2p$ electrons bound by 54.9~eV.
There are three continuum electrons
inside the WS sphere $R_\text{\tiny WS}=2.99$~(a.u.).
The continuum density inside the sphere $n_c(r)$ converges to the uniform
free-electron density $n_e = Z_f /V_{\text{\tiny{WK}}}$,
where $Z_f = 2.146$.  In Fig.~\ref{al2}, the structure function $S(k,\omega)$
is plotted for the case of an incident 2.96~keV photon scattered at 30$^\circ$.
The coherence parameter $\alpha = 1.95$ in this case, explaining the prominent
plasmon resonance seen on the low-frequency side of the elastic scattering peak.
Also shown in the figure are contributions from the bound $L$ shell electrons
scaled up by a factor of 100.
 It should be noted that,  by contrast with the average-atom
calculations presented in Ref.~\cite{SG:08}, the $M$ shell of Al is completely
empty at temperatures below 10~eV in the present model and the
prominent $M$ shell
features predicted in Ref.~\cite{SG:08} do not arise in the present analysis.

\subsection{Titanium at $T =10$~eV and metallic density}

\begin{figure}[t]
\centerline{\includegraphics[scale=0.7]{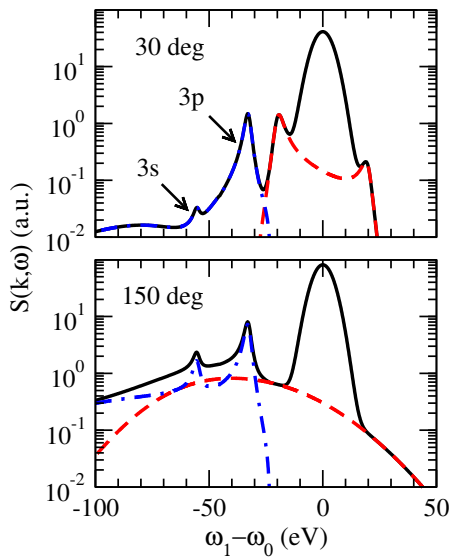}}  
\caption{(Color online)  Dynamic structure function $S(k,\omega)$ for
scattering of a 2960~eV photon at 30$^\circ$ and 150$^\circ$ from
metallic density Ti
at $T_e=10$~eV. Contributions to $S(k,\omega)$ (solid curve)
from $S_{ee}(k,\omega)$ (dashed curve) and  $S_\text{\tiny B}(k,\omega)$
(dot-dashed curve) are shown.
\label{ti10}}
\end{figure}

Titanium ($Z=22$) at metallic density ($\rho = 4.51$~g/cc) and $T = 10$~eV
is a case where sharp peaks from excitations of bound $M$ shell electrons
show up in the energy spectrum along with the plasmon peaks.
The average-atom model predicts that metallic density Ti
is in an Ar-like configuration
at $T=10$~eV with completely filled K and L shells together with 1.97 $3s$
electrons bound by 44.40~eV and 5.36  $3p$ electrons bound by 22.88~eV.
There are 4.67 continuum electrons inside the WS sphere
$R_\text{\tiny WS}=3.05$~(a.u.).
The continuum density inside the sphere converges to the uniform
free-electron density $n_e = Z_f/V_{\text{\tiny{WK}}}$ outside the sphere,
with $Z_f = 2.305$. The chemical potential is $\mu=-0.0511$~au.
In Fig.~\ref{ti10}, the dynamic structure function
$S(k,\omega)$ is shown for the case of an incident 2.96~keV photon scattered
at 30$^\circ$ and 150$^\circ$. Plasmon peaks, which are prominent for
scattering at 30$^\circ$, disappear for scattering at 150$^\circ$
while the $3s$ and $3p$ bound-state peaks grow. The $M$ shell contributions
to the structure function are comparable to the plasmon contribution
for the 30$^\circ$ case and are the dominant features on the low-frequency
side of the elastic peak at 150$^\circ$.


\section{Summary}

A scheme for analysis of Thomson scattering from plasmas based
on the average-atom model, a quantum-mechanical version of the
``Generalized Thomas-Fermi Theory" of
Feynman, Metropolis and Teller \cite{FMT:49} is presented.
Given the plasma composition $(Z,A)$, density $\rho$ and temperature $T$,
the model gives, in addition to the equation of state of the plasma,
all parameters needed for a complete description of the Thomson scattering
process. In particular, the average-atom code predicts wave functions
for bound and continuum electrons, densities of bound, screening and
free electrons, and the chemical potential.

Predictions of the present average-atom model disagree with those in
Ref.~\cite{SG:08} where a similar model with different boundary conditions is used.
In particular, in Ref.~\cite{SG:08}, $3d$ electrons were bound in
metallic density Al
for temperatures between 2 and 10~eV, leading to substantial bound-state
contributions to the dynamic structure function. In the present model
the $3d$ subshell of metallic density Al is vacant in the temperature
range $T\leq 10$~eV and the corresponding bound-state features are absent.

Elastic scattering from bound and screening electrons is treated following
the model proposed by \citet{GG:03} which makes use of formulas for the
static ion-ion structure function $S_{ii}(k)$ given by \citet{AD:98}.
Modifications suggested by \citet{GGL:06} to account for different electron
and ion temperatures are also included. Treatment of the ion-ion structure
function appears to be the weakest aspect of the present analysis.
The dynamic structure function for scattering from free electrons depends
sensitively on the free-electron dielectric function $\epsilon(k,\omega)$.
Again, we follow the model proposed in Ref.~\cite{GG:03} and evaluate
the dielectric function in the random-phase approximation. The RPA dielectric
function includes features such as plasmon resonant peaks that show up in
experimental intensity profiles and can be used in connection
with the principle of detailed balance to determine electron temperatures.
Bound-state features are included in the present scheme, inasmuch as the
average-atom model provides bound-state and continuum wave functions.
Coulomb-field effects are automatically included in calculations
carried out using average-atom continuum states rather than plane waves
to describe the final state electron.
In conclusion, the average-atom model provides a simple and consistent point
of departure for theoretical analysis of Thomson scattering from plasmas.

\section*{Acknowledgements}
We owe debts of gratitude to S. H. Glenzer, C. Fortmann and T. D\"{o}ppner
for informative discussions and for providing comparison experimental data
for x-ray scattering from beryllium.
The work of J.N.\ and K.T.C.\ was performed under the auspices of the U.S.\
Department of Energy by Lawrence Livermore National Laboratory
under Contract DE-AC52-07NA27344.


\end{document}